\documentclass[review]{elsarticle}
\usepackage{amsmath}
\usepackage{cleveref}
\usepackage{xspace}
\usepackage{bold-extra}
\usepackage{amsfonts}
\usepackage{amsthm} % definte the proof environment: block letters + italic letters
\theoremstyle{definition}

\usepackage{comment}
\usepackage{xcolor}
\usepackage{subcaption}
\usepackage{epstopdf}% To incorporate .eps illustrations using PDFLaTeX, etc.
\begin{document}
\begin{frontmatter}

%\articletype{RESEARCH ARTICLE}% Specify the article type or omit as appropriate

\title{Stochastic physics of species extinctions in a large population}
%Species extinctions under stochastic environmental forcing

\author[rvt]{Ivan Sudakov\corref{cor1}}
\ead{isudakov1@udayton.edu}
\author[focal,els]{Sergey A. Vakulenko} 
%\ead{vakulenfr@gmail.com}
\author[rvt1,rtv2]{John T. Bruun} 
%\ead{dkirievskaya1@udayton.edu}

\cortext[cor1]{Corresponding author}
\address[rvt]{Department of Physics, University of Dayton, Dayton, OH 45469-2314, USA}
\address[focal]{Institute of Problems in Mechanical Engineering, Russian Academy of Sciences, St.\,Petersburg 199178, Russia}
\address[els]{Department of Electrical Engineering and Precision Electro-Mechanical Systems, ITMO University, St.\,Petersburg 197101, Russia}
\address[rvt1]{College of Engineering, Mathematics and Physical Sciences, University of Exeter, Exeter EX4 4QJ, UK}
\address[rtv2]{College of Life and Environmental Sciences, University of Exeter, Penryn Campus, Penryn TR10 9FE, UK.}

\begin{abstract}
Species extinction is a core process that affects the diversity of life on Earth. Competition between species in a population is considered by ecological niche-based theories as a key factor leading to different severity of species extinctions. There are population dynamics models that describe a simple and easily understandable mechanism for resource competition. However, these models can not efficiently characterize and quantify new emergent extinctions in a large population appearing due to environmental forcing. To address this issue we develop a stochastic physics-inspired approach to analyze how environmental forcing influences the severity of species extinctions in such models. This approach is based on the large deviations theory of stochastic processes (the Freidlin-Wentzell theory).  We show that there are three possible fundamentally different scenarios of extinctions, which we call catastrophic extinctions, asymmetric ones, and extinctions with exponentially small probabilities. The realization of those scenarios depends on environmental noise properties and the boundaries of niches, which define the domain, where species survive. Furthermore, we describe a hysteresis effect in species extinction showing that fluctuations can lead to dramatic consequences even if an averaged resource supply is sufficient to support population survival. Our stochastic physics-inspired approach generalizes niche theory by accounting environmental forcing and will be useful to find, by available data, which environmental perturbations may induce extinctions.

\end{abstract}


%\subjclass{Primary 99Z99; Secondary 00A00}

\end{frontmatter}

%\linenumbers

\section{Introduction} \label{Intro}

%In order to assist authors in the process of preparing a manuscript for a journal, the Taylor \& Francis `\textsf{Interact}' layout style has been implemented as a \LaTeXe\ class file based on the \texttt{article} document class. A \textsc{Bib}\TeX\ bibliography style file and a sample bibliography are also provided in order to assist with the formatting of your references.

%Commands that differ from or are provided in addition to standard \LaTeXe\ are described in this document, which is \emph{not} a substitute for a \LaTeXe\ tutorial.

%The \texttt{interacttfssample.tex} file can be used as a template for a manuscript by cutting, pasting, inserting and deleting text as appropriate, using the preamble and the \LaTeX\ environments provided (e.g.\ \verb"\begin{abstract}", \verb"\begin{keywords}").

Life on earth has co-evolved with fluctuations in both climate and the wider abiotic environment. However, the most pronounced changes in biota come during periods of extreme environmental perturbation. These are commonly associated with cataclysmic events (e.g. bolide impact) \citep{bolid16}, rapid shifts in the climate state \citep{ocean17} or a combination of the the both factors \citep{jo17}. During these periods the biotic and abiotic changes feedback on each other.
Generally, the two-way interactions between populations in ecosystems and their abiotic characteristics have the potential to dramatically impact both biodiversity \cite{Smyth,Huisman2006,Tarran2015,Cornwell2018,Cornwell2020}, and a wider planetary response to the changing environment \cite{Arr12,abby}.

Mass extinctions represent important effects that correspond to biodiversity loss, ecosystem upheavals and changes to the evolution of life.
%Modern biodiversity loss and the possibility of it representing a 6th mass extinction caused by human pressures, makes it pertinent to understand the dynamics of past mass extinctions. 
This is seen in the fossil record through pronounced
changes in fossil assemblages with mass extinctions being attributed to large-scale environmental disasters \citep{Jab05}. However, their
dynamics are often poorly understood: it is often unclear as to why populations decline to the point of a species becoming extinct. Here we assess a generic range of environmental forcing types and its impact on the dynamics of species extinction and evolution. These forcing characteristics are  macroscopic and emergent  properties  of small scale wave interaction processes for the ocean - atmosphere - land system.

Some studies \cite{Smyth,Tarran2015,Cornwell2018} show that extinctions play a major role in maintaining population dynamics for niche species due to the niche-overlap in growing species populations increases and co-existence becomes more uncertain \cite{Chesson2000}. Employing stochastic physics is a perspective way to develop a better understanding of how the scenarios of extinction depend on environmental fluctuation features and the boundaries of niches. In this paper,  we develop an approach based upon recent studies investigating the connection between ecological dynamics in a  changing environment and the statistical/stochastic physics of large systems  \cite{Reg89,Tok04, Mehta, Kes15, Dick16, Tix}. However, in contrast to these previous works,  we propose to study the multistability effect in a system with a large population and fixed parameters replacing it to a similar system with slowly evolving parameters and observable jumps between equilibria. This technique is originally proposed in the large deviations theory of stochastic processes and has been introduced in the work of M. Freidlin and A. Wentzell \cite{FW}.

We propose a general s stochastic population competition model that allows the assessment of how a wide range of environmental forcing types can affect biodiversity (the number of coexisting species), biomass (the number of species in a population) and extinction. 
%The stochastic dynamics of this model are strongly dependent on initial parameters. 
In that model, extinctions are inevitable if a population has the maximal possible biodiversity and uses the maximal amount of resources.
%Our modeling provides a quantitative framework in which to investigate the dynamics of biospheric change.  Using analytical investigations and numerical simulations we show that the presence non-linear process effects can be a key component for extinction events. %Thus, theoretical investigation of the dynamical properties of large ecosystems under external forcing makes a significant contribution to understanding properties of the ecosystems as well as the climate system.The large ecosystem is the open systems which may stay under external forcing. 
Also, we use fairly general assumptions. First, each species can survive within  a niche, so-called  Hutchinsonian niche in environmental parameter space \cite{Hutch}. That niche is an "n-dimensional hypervolume", where the dimensions are environmental conditions and resources. Furthermore, dynamics of environmental forcing is defined by stochastic dynamical systems with a small noise. Then 
%by basic results of the Freidlin-Wentzell theory \cite{FW}
we show that there exist three different regimes of extinctions: all species become extinct in a very fast way, all species become extinct very slowly, or part of the species become extinct quickly with the remaining ones slowly. The occurrence of the situations depends on mutual location of the system attractors that define the environmental forcing type and species niches.

We use a general range of environmental forcing types: time quasiperiodic oscillations, a simple white noise, and chaotic dynamics with many attractors and a weak noise. We assume that such changes in the environment may influence the states of the system of species. Indeed, in the Earth's climate system there is the El Ni\~{n}o-Southern Oscillation (ENSO), short term ($3$ to $10$ year) cycles of nonlinear interaction of wind with the equatorial waveguide in the Pacific. There are many shreds of evidence that this phenomenon affects the ecosystems as a form of chaotic forcing effect~\cite{Bruun}. These chaotic and stochastic effects can occur due to turbulence and small scale wave interaction processes in the surface wind stress and internal ocean and that these can play an important role in climate \cite{Williams2012, Bruun2019}. When a system exhibits periodic forms of chaos, the stochastic nature of the system leads to eigenmode repulsion: the spacing between adjacent eigenmodes follows a universal Gaussian ensemble and Random Matrix Theory (RMT)~\cite{Bruun2019} is used to identify these stochastic system types by assessing emergent resonance phenomena. It helps to reveal how internal system wave mechanisms interact and so influence the ecosystems in their vicinity.

%The species model exhibits a highly stochastic behavior. The final population state depends dramatically on initial data. For some initial abundances all species coexist, whereas for other initial data only a few species can survive for large times. Usually, the environment influence diminishes the number of surviving species. Systems with large numbers of species are more stable than the ones with few species. The multistability effect in a system with fixed parameters means that in a similar  system with slowly evolving parameters we can observe jumps between equilibria and this process is investigated using the Freidlin-Wentzell method \cite{FW}.

The paper is organized as follows. In Section \ref{forcing} we state the niche model of species coexistence and we describe different models of environment fluctuations. In Section \ref{newnoise} we explain our main result:
the description of  general different extinction scenarios. 
In this section the Freidlin-Wentzell theory \cite{FW} is applied to describe influences of a weak noise (which is further outlined in the Appendix). In Section \ref{SM} we study the extended dynamical model of a population competing for resources \cite{HuWe99, Vak16, SVGK, Vak17}, which takes into account species extinctions and time oscillations of the resource.
This model is an extension of the well-known J. Huisman and F. Weissing model \citep{HuWe99} that has been used to study phytoplankton. The model accounts for species self-regulation, extinctions, and time dependence of resources. 
For large resource turnovers  this model has a simple asymptotic solution. Section \ref{hysteff} discusses the qualitative processes of the extinction in a population dynamics model. In Section \ref{numres}  we describe the results of numerical simulations.

\section{Niche model and environmental fluctuations} \label{forcing}
%plant or marine (plankton)
We consider a system of species that depend
on the environmental state $q$. For example, plants or plankton species
depend on a few of resources, and certain resources are directly connect with the environment (sun light, temperature, $CO_2$ concentration etc.).  
To do this we utilize the ideas of niche theory \cite{Mehta}: a $j$-th species survives only when environment parameters lies within a
given domain denoted by $\Omega_j$.  

Let us suppose that a system occupies a specific area or region. We denote the averaging over that area as well as the environmental parameters (which are essential for species  survival) by $(q_1, q_2,..., q_n)=q \in {\mathbb R}^n$. This property can also be time $t$ dependent. The key specification for niche survival is that the $j$-th species survives
while $q(t) \in \Omega_j$. Each domain (niche) $\Omega_j$ is a bounded subset of 
${\mathbb R}^n$ with a smooth boundary $\partial \Omega_j$. In this model, extinctions occur when $q$ goes through the boundary
 $\partial \Omega_j$ for certain $j$ and leaves the domain $\Omega_j$. We define a mass extinction as occurring when $q$ leaves a number of the domains $\Omega_j$
 within a specific time period.

%The time $T_s$ when at least one species 
%with $x_j(t) >X_{ext}$ exists is called the survival time.   
%In this section, we also introduce the models of chaotic and periodic 

The generic form of environmental forcing $q$ reveal how internal system wave mechanisms interact and so influence species in their vicinity. These environmental forcing and species response impacts range from oscillatory, chaotic and noisy-stochastic physical perturbations. Our specification here accommodates these entire range of perturbation types. Periodic and quasiperiodic environmental oscillations and stochastic effects can also also impact the extreme range leading to perturbations in species which can subsequently exhibit this a non-linear and chaotic responses \cite{Bruun, Williams2012, Bruun2019,Shaffer1989}. 
%In CGM studies stochastic simulation schemes are used to represent these forms of fluctuation.
When a system exhibits periodic forms of chaos, the fundamental random and stochastic nature of the system leads to eigenmode repulsion: the spacing between adjacent eigenmodes follows a universal Gaussian ensemble. In that setting including RMT provides the ability to assess emergent species survival 
properties \cite{Bruun2019}.

In all cases, we suppose
that the system depends on environment state via the resource supply
\begin{equation} \label{bartildS}
    S=\bar S + \tilde S(q(\tau)), 
\end{equation}
where the perturbation $\tilde S$ of the background resource supply
$\bar S$ 
and dynamics of the environment state $q$ is slow, i.e., $\tau =\kappa t$, $\kappa <<1$. In this work we assess %(using methods of the Freidlin-Wentzell theory, see \cite{FW} and Appendix)
scenarios of extinction under noise. This includes the range of environmental forcing types given below, which provide a range of forcing types from oscillatory motion, noisy systems to a non-linear chaotic forcing type that accommodates all the oscillatory and noisy settings.
%To simulate forcing we set 
%\begin{equation} \label{supplychaos} 
  % S=S_0 +  r \theta(q(t)) 
%\end{equation}
 
 \subsection{Periodic and quasiperiodic variation} \label{quasiper}
 
 The simplest model, that simulates the effect induced by 
 large scale environmental forcing scenarios \cite{Bruun}, is
 \begin{equation} \label{quasi}
     q(\tau)=q^0 +  Re \ \sum_{j=1}^m c_j \exp(i \omega_j \tau), 
 \end{equation}
 where $q^0$ is a constant, $i=\sqrt{-1}$, $c_j$ are coefficients and
 $\omega_j$ are frequencies (periodic and/or quasiperiodic oscillations). Model (\ref{quasi}) can be extended to include a very slow variation or trend as 
 \begin{equation} \label{quasitrend}
     q(\tau)=q^0 + \alpha \tau +  Re \ \sum_{j=1}^m c_j \exp(i \omega_j \tau).
 \end{equation}
 
 \subsection{Dynamical forcing with noise} \label{DF}
Where the dynamics of $q$ is governed by trajectories of a noisy dynamical system, we write this in the Ito form
\begin{equation} \label{dynsysST} 
   dq=Q(q)d\tau  +  \sqrt{\epsilon}  \ dB(\tau),
\end{equation}
where $B(t)$ is the standard Brownian motion and $Q$ is a smooth vector field,
$\epsilon >0$. In the case $\epsilon=0$ the system explanation reduces to the differential equation
\begin{equation} \label{dynsys}
\frac{dq}{dt}=Q(q), 
\end{equation}
and we suppose that its dynamics are well posed and this has a compact attractor ${\mathcal A}_Q$. 
By varying different $Q$, we can obtain different kinds of noise induced forcing. The emergent RMT properties are a feature of such noise induced systems. In this noisy type where small $\epsilon>0$ we can also apply the Freidlin-Wentzell theory \cite{FW} and also see the Appendix.

\subsection{Bistable state transitions} \label{Q3}

\begin{figure}[t]
\includegraphics[width=0.7\linewidth]{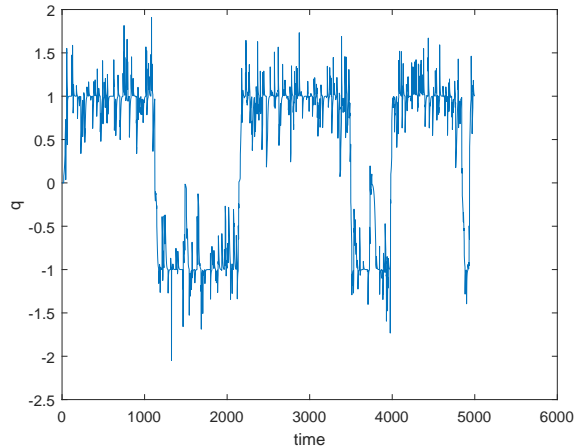} 
\caption{This plot shows Kramers transitions for
model from subsection \ref{Q3}. We set $Q=q-q^3$, $q \in {\bf R}$ and the noise $B(t)$ is a sequence of random outliers. For each time $t$ we have either $B(t)=r$ with probability $p_{out}$ (where $r$ is a random number) or otherwise
$B(t)=0$. We sample $r$ from the normal distribution, $r \in {\bf N}(0, 3)$.} 
\label{Kramer}
\end{figure}

The generic model we select to represent bistability is obtained by
setting $q \in {\bf R}$ and $Q=aq -q^3$, where $a$ is a parameter.
For $a >0$ the attractor consists of two stable points, $q= \pm a^{-1/2}$. In noisy systems, this system exhibits random 
transitions from state $q=1$ to state $-1$ and back. By varying the size of $\epsilon$ the occurrence of such transitions can be regulated, for small (large) $\epsilon$ these transitions are rare (frequent), see Figure \ref{Kramer}. 

\subsection{Chaotic forcing} \label{ChF}

%and consider the Lorenz system, a rough model of atmospheric dynamics given by 
%\begin{equation}
%\label{A.12} 
%\begin{split} 
%dx/d\tau=\alpha(y-x)    ,\\
%dy/d\tau=x(\rho -z) -y  ,\\
%dz/d\tau=xy - \beta z ,
%\end{split}
% \end{equation}
%where $\alpha, \beta, \rho$ are parameters.
%For $\epsilon=0$ that system shows a chaotic behaviour for $\alpha=10, \beta=8/3$ and $\rho=28$.
%We assign the function $\theta$ as follows. The third component $z$ in (\ref{A.12}) describes the time evolution of temperature $T=z$. 
 
A generic representation of chaotic forcing is the non-linear delay oscillator. This model is
\begin{equation} \label{ENSO}
    \frac{dq}{d\tau}=a F(q(\tau-\tau_1)) - a F(q(\tau-\tau_2)) + c\cos(\omega \tau),
\end{equation}
where $a, b, c, \tau_k$ are positive and non-linearity 
$F$ can be chosen, for example, in the form $F(q)=q-q^3$. This accommodates
regular seasonal forcing, non-linear wave interaction processes and time delays and it is discussed in detail in \cite{Bruun}. This type of forcing generically corresponds to a periodic map, with universal properties for this established by Feigenbaum  \cite{Feigenbaum1980}. It is a model for cascades into turbulence and sub-harmonic resonance that applies to all such periodic map processes. This chaotic forcing exhibits high frequency intermittency \cite{Bruun} and slow variation modes consistent with centennial time scales \cite{Skakala2018}. This dynamic model has a wide range of properties and used to represent the generic wave-interaction ENSO process. It can be used to move through states of oscillatory motion, bifurcations, chaos and intermittency. Periodic map properties also occur for species \cite{Huisman2006}.

%Before we present some numerical results, we will 
%Here we assess (using methods of the Freidlin-Wentzell theory, see \cite{FW} and Appendix) scenarios of extinctions under noise.

\section{The scenarios of extinction under environmental forcing} \label{newnoise}

Our primary goal  is to find the probabilities of extinctions in our model. 
We consider three sharply different extinction scenarios of this which can be generated by random and non-random environment forcing induced by equation \eqref{dynsys}. Using the well known results \cite{FW} (see Appendix) we establish that there are three possible extinction  scenarios as a function of the noise magnitude $\epsilon$ and mutual locations of the sets  ${\mathcal A}_Q$ and  $\partial \Omega_j$. Let us remind that  ${\mathcal A}_Q$  denotes an attractor of dynamical system (\ref{dynsysST}) for $\epsilon=0$ and $\partial \Omega_j$ denotes a boundary of existence of $j$-th species in the space parameter. In our model , that boundary is defined by the resource supply $S=S(q)$, where $q$ evolves according to (\ref{dynsysST}). Let us denote by $P_{j, \epsilon}$ the probability of extinction of the $j$-th species per a fixed time period (here $\epsilon$ is the noise level, see the previous section).

By arguments stated in Appendix  we find  the following three sharply different scenarios of extinction types:
\\~\\
{\bf I. Catastrophic species extinctions}: If the intersection $I={\mathcal A}_Q \cap {\partial \Omega} _j $ is not empty for all 
 $j=1, ..., k$ then 
the probability $P_{j,\epsilon}$ is not exponentially small, i.e., 
$\lim_{\epsilon \to 0} \epsilon \log P_{j, \epsilon}>0$. It is a catastrophic scenario when the extinction of all species  (mass extinction) is quite probable.
\\~\\
{\bf II. Species extinctions with exponentially small probabilities}:
The intersection $I={\mathcal A}_Q \cap  {\partial \Omega}_j$ is empty for all $j$.
   Then 
the probabilities $P_{j,\epsilon}$ are  exponentially small both for large and small extinctions. 
\\~\\
{\bf III. Asymmetric species extinctions}:
The intersection $I={\mathcal A}_Q \cap {\partial \Omega}_j$ is not empty  for  some
$j$ but  it is  empty for others $j$. Then it is possible that
the probability $P_{j,\epsilon}$ is not small for extinctions involving relatively few species but that probability is exponentially small for extinctions involving relatively many species. In this case, there is a sharp transition in the probabilities of small losses of biodiversity and great losses of biodiversity. 
\\~\\
When the attractor consists of $n_A >1$ connected components ${\mathcal A}_Q^{(i)}$ we find that there are possible additional effects that may be caused by bifurcations in the environment system. For example, some climate models exhibit a possibility of climate  bifurcations (tipping points) \citep{Len11,Sud13} with rapid changes of the climate system from one stable state to another. With the non-linear delay oscillator dynamic (\ref{quasitrend}) and (\ref{ENSO})  transitions from rapidly varying intermittent to slowly modulated cyclic forcing can be accommodated \cite{Bruun}. The species resilience impact  of these three extinction types may correspond to a variation between system state \cite{Huisman2006}. One can suppose that the climate bifurcations  may be caused by a transition from a connected component to another one: for example, by a transition from scenario I to scenario II (or III), and vice versa.
Overall we can say that extinctions in such models are completely predetermined by the distances between the local attractors of the noisy dynamical systems that generates the environmental forcing  and critical resource level sets. It is worth noting that among the variants
considered in subsection \ref{ChF}
 the most interesting case is defined by the equation (\ref{ENSO}), which provides a framing to encapsulate all the scenarios of interest.

\section{The resource model} \label{SM}

We consider the following standard model of biodiversity~\cite{HuWe99}:
\begin{equation}
     \frac{dx_i}{dt}=x_i (- r_i  + \rho_i(v) -  \sum_{j=1}^N \gamma_{ij} \; x_j),
    \label{HX11}
     \end{equation}
\begin{equation}
     \frac{dv}{dt}=D(S-v)   -  \sum_{j=1}^N c_j \; x_j \; \rho_j(v),
    \label{HV11}
     \end{equation}
where
\begin{equation}
      \rho_j(v)= \frac{a_j v}{K_j +v}, \quad  a_j , \ K_j >0,
\label{MM5}
     \end{equation}
are Michaelis-Menten's functions,  $x_i$ are species abundances,
$r_i$ are the species mortalities, $D$ is the resource turnover rate, 
$S$ is the supply of resource $v$, and
$c_i$ is the content of the resource 
in the $i$-th species. These constants
define how different species share resources. 
Note that if $c_i=0$ then
the equation for $v$ becomes trivial and 
$v(t) \to S$ for large times $t$, i.e., the resource
equals the resource supply. 
The terms $\gamma_{ii}x_i$ define
self-regulation of species populations that
restrict the species abundances, 
and $\gamma_{ij}x_j$ with $i \ne j$ define a possible competition  between species for resources. 
The coefficients $a_i$ are 
specific growth rates and the $K_i$ are self-saturation constants. 
If $\gamma_{ij}=0$ this system is equivalent to those in works 
where the plankton paradox~\cite{Hof88} is studied. For the case of $M$ resources we have   more complicated equations
\begin{equation}
     \frac{dx_i}{dt}=x_i (- r_i  + \phi_i(v) -  \sum_{j=1}^N \gamma_{ij} \; x_j),
    \label{HX2}
     \end{equation}
\begin{equation}
     \frac{dv_j}{dt}=D_j(S_j -v_j)   -  \sum_{k=1}^N c_{jk} \; x_k  \; \phi_k(v),
    \label{HV2}
     \end{equation}
where $v=(v_1, v_2, ..., v_M)$,   and
\begin{equation}
      \phi_j(v)= \min \{ \frac{a_j v_1}{K_{1j} + v_1}, ...,  \frac{a_j v_M}{K_{Mj} + v_M}  \} \, .
\label{MM2}
     \end{equation}
with $a_j$ and $K_{ij} >0$.
This model is widely used for primary producers like phytoplankton and it can also 
be applied to describe competition for terrestrial plants 
\cite{Til77}. Relation (\ref{MM2}) corresponds to the von Liebig minimum law, but we can consider even
more general $\phi_j$  satisfying  the conditions
\begin{equation}
      \phi_j(v) \in C^1, \quad    0 \le \phi_j(v) \le  C_+,  
\label{MM2a}
     \end{equation}
where $C_{+} >0$ is a positive  constant, and
\begin{equation}
      \phi_k(v) =0,   \quad  \forall k \quad v \in \partial {\bf R}^N_{>}     
\label{MM2b}
     \end{equation}
where $\partial {\bf R}^N_{>} $ denotes the boundary of the positive cone $ {\bf R}^N_{>} =\{v:  v_j \ge 0, \ \forall j\}$.
Note that condition (\ref{MM2b}) holds if $\phi_j$ are defined by   (\ref{MM2}).
%Moreover, we assume that $\phi_j(z)$ are increasing 
Similarly as above, we  assume that
$
\sum_{k=1}^N c_{ik}=1, \quad  c_{ik} >0.
$
This model is well posed. Under certain natural 
conditions to $\gamma_{ij}$ solutions are defined for all positive
times $t$, they are unique and there exists a finite dimensional attractor
\cite{Vak16}.

\section{The hysteresis effect}
\label{hysteff}

Suppose the resource supply $S$  depends on an external parameter,  for example, temperature,
$T$, which evolves very slowly through a long term Ocean basin scale mode \cite{Bruun,Skakala2018} or similar slowly varying environmental process. Bruun et al. \cite{Bruun} showed that at a Pacific basin scale the tree-growth and temperature hysteresis exhibited persistent cycles that contributed to epochs of excessive heat and cold over a seven century time period. That work showed that the cyclic modes became unstable at the extreme range of the hysteresis curve, however  those  modes stayed stable overall, and indicated that the current warming period is at the hysteresis upper thermal edge for environmental forcing dynamics. 

\begin{figure}[t]
\includegraphics[width=0.6\linewidth]{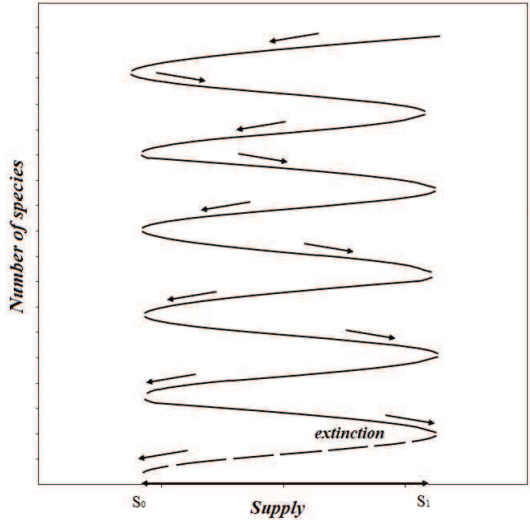} 
\caption{The representation of an extinction process. Each cycle of change of resources supply from $S_1$ to $S_0$ and back leads to a decrease in species number $N$. As a result, a population may become extinct.} \label{hysty}
\end{figure}

It is natural to suppose  that at each $T$ our system in an equilibrium state.  Let $S(T)$ be an increasing function of $T$. Then 
as $T$ increases  from $T_0$ up to $T_1 >T_0$ we can observe a bifurcation sequence described in the previous subsection. In our model we obtain that our system is, in sense, invertible, i.e., as $T$ changes from $T_1$ to $T_0$, we observe the same bifurcations but going in the reverse order.  Thus, in our ideal model slow environmental oscillations do not affect biodiversity.
However, in a more realistic situation, where  species extinct if their abundance is less than a certain
threshold $X_{ext}$ (this model is considered in \cite{Vak16} in another context), then 
 environmental oscillations can lead to partial loss of species. In our model by letting $T$ slowly go from
$T_1$ to $T_0 < T_1$ so that for the abundance of a species at $T=T_0$ falls beneath
the threshold, then this species never returns in our system, even when $T$  returns to the start value $T=T_1$. This effect is  illustrated by the Figure \ref{hysty}. 

%following pictures done by numerical simulations for the standard model with a single resource $v$.

\begin{figure}[t]
\includegraphics[width=0.7\linewidth]{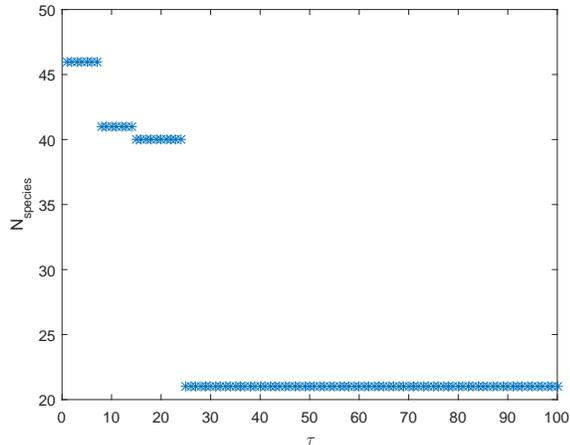} \label{Fig3H}
\caption{This plot shows as noisy environmental forcing can lead to extinctions. Here we do not take into account hysteresis, i.e., we suppose that population restores after each environmental
shock (the time like parameter $\tau$ numerates shocks). We consider an equilibrium state for standard model (\ref{HX2}, \ref{HV2}) with a single resource and the number of species $N_{species}$ coexisting in that equilibrium state. The model parameters are:
initial species number $N_{pool}=50$, $b_i=2$, $K_i=4$, where $i=1, ..., N_{pool}$, $D=5$, 
and $S=10$.  The mortalities $r_i$ are independent random numbers distributed normally according to ${\bf N}(1, 0.1)$. The matrix $\gamma_{ij}$
is diagonal with entries $\gamma_{ii}=1$. The noise is induced
by random outliers of $S$ at times $t=\tau, 2\tau$ ...
where $\tau>>1$. These outliers are defined by 
so $S(\tau)=\bar S + \tilde S(\tau)$, where $\tilde S(\tau)$
are independent random numbers normally distributed according to 
${\bf N}(0, 3)$ and $\bar S=10$.
} \label{noise}
\end{figure}

\begin{figure}[t]
\includegraphics[width=0.7\linewidth]{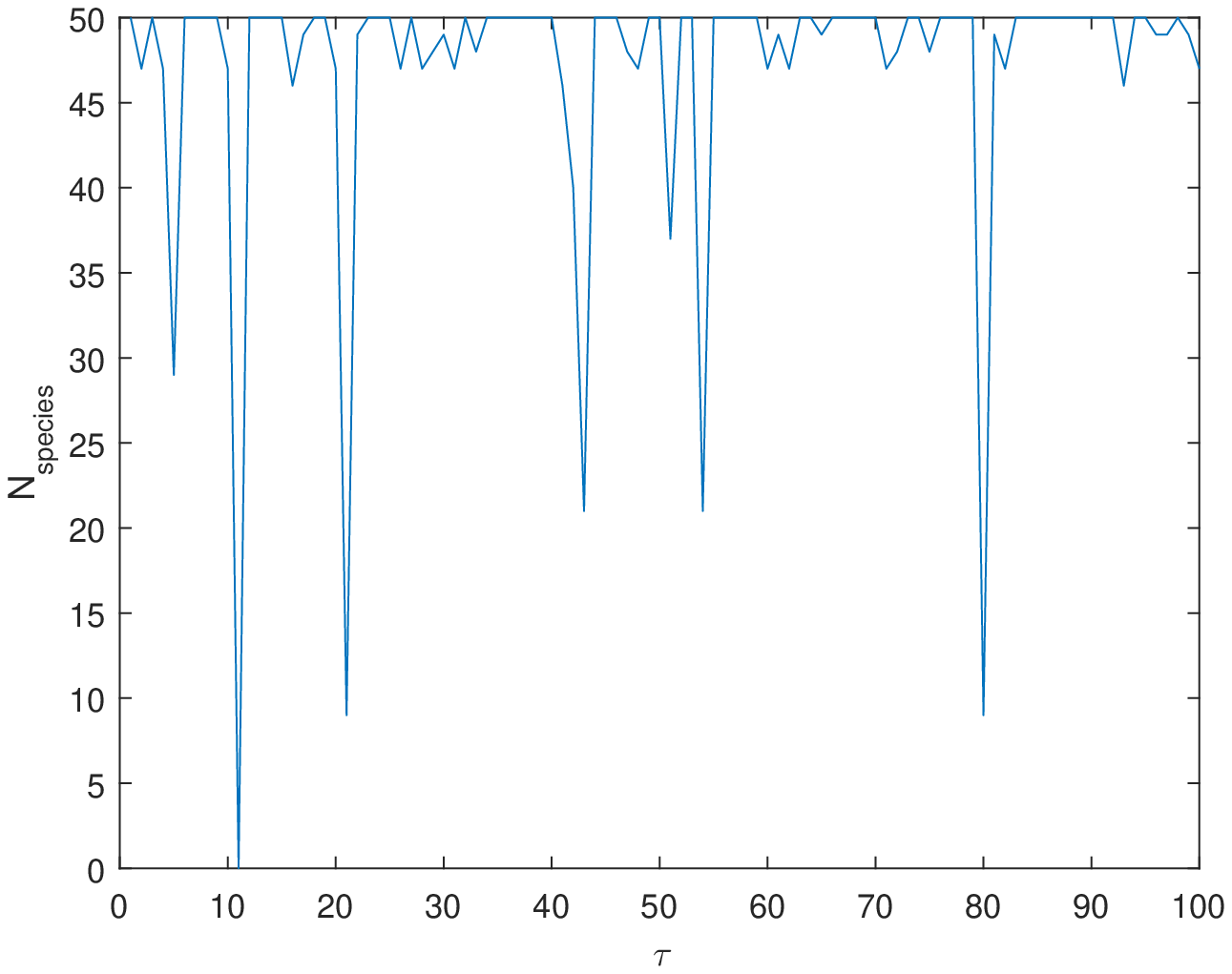} 
\caption{Noisy environmental forcing leading to extinctions. Here we take into account hysteresis, i.e., we suppose that population does not restore after each environmental
shock. We consider an equilibrium state for the standard model with a single resource and the same parameters as on the previous image.
%The noise also is as above.
}
\label{noisy2}
\end{figure}

\subsection{Analytical study of hysteresis in noisy environment}

To investigate hysteresis effect, we consider the simplest case
of model (\ref{HX11},\ref{HV11}) with a diagonal matrix
$\gamma_{ij}=\delta_{ij} \gamma$, identical parameters
$K_i=K$ and $b_i=b$ and random mortalities $r_i$, which are random positive numbers distributed according to a density
$\rho(r)$. 
Moreover, let $c_{k}=c>0$. Then it is natural to introduce
notation $\phi(v)=b v/(K +v)$.
For fixed  $S$ and large $D$  the system is in an equilibrium state defined by  \cite{Vak16}
\begin{equation} \label{eqstatev}
D(S- v_{eq}) = c B_M  \; \phi(v), 
\end{equation}
where
$$
B_M=\sum_{k=1}^N X_k(v_{eq})
$$
is a biomass of the system and 
$$
X_k=  \gamma^{-1}(\phi(v_{eq}) - r_k)_{+} 
$$
are steady state species abundances and $f_+=\max \{0, f\}$.

We consider the following simple model of noise in $S$.
 Suppose  at certain moments $\tau_j=j \Delta t$ we have jumps in $S$: $S(\tau_j+0)=S(\tau_j+0) +\Delta S_j$, those outliers $\Delta S_j$ can have different signs. Moreover, we assume that
 the interval $\Delta t$ is much more than the characteristic relaxation time $t_{eq}$, thus most of the time within the  interval $I_j=[\tau_j , \tau_{j+1})$ 
  system (\ref{HX11}), (\ref{HV11}) is an equilibrium state
  corresponding to the resource supply value $S(\tau_j+0)=S + \Delta S_j$. At initial time moment we have a pool of species with $N=N_{pool}>> 1$ of species. 
 Let $N_{sp}(\tau_j)$ be the number of species with non-zero abundances at $t \in (\tau_j, \tau_{j+1})$, i.e, biodiversity within the interval $I_j$.
Then for the number $N_{sp}$ one has the following recurrent 
relation:
$$
N_{sp}(\tau_{j+1})=N_{sp}(\tau_{j})\mu_j
$$
where $\mu_j$ is a fraction of species, which  survive after  $j$-th outlier. For $N_{sp}>>1$ this fraction can 
be estimated as follows. Note that if $k$-th species survives,
i.e. $X_k(v_{eq}) >0$ then  $r_k > \phi(v_{eq})$. The value
$v_{eq}$ increases in $S$. Let us denote by
$V_j$ the equilibrium value $v_{eq}$ for
$t \in [\tau_{j-1}, \tau_j)$.
For $\Delta S_j \ge 0$  one obtains
that $\mu_j=1$. For $\Delta S_j < 0$ one has
$$
\mu_j  \approx  \frac{\int_{-\infty}^{\phi(V_{j+1})} \rho(r) dr }{
\int_{-\infty}^{\phi(V_{j})} \rho(r) dr}
$$
  Consider the sequence of $S(\tau_j)$, $j=1,..., N_S$. This sequence can be decomposed into increasing and decreasing sub-sequences.
  Similarly, the corresponding sequence of equilibrium values
  $V_j$ falls into analogous increasing and decreasing sub-sequences since $v_{eq}(S)$ is a monotone increasing in $S$
  function.
  As $S$ non-decreases, the population diversity conserves
  and all species survive. Thus increasing intervals change nothing in diversity. Consider 
  a decreasing interval, which starts with
 $S=S_b$, and finishes at a local minimum of $S$, which equals
 $S_e$. 
 A change in species diversity within such decreasing interval is defined then by 
 $$
 \frac{{N_{sp}}^{(e)}} {{N_{sp}}^{(b)}}\approx \frac{\int_{-\infty}^{\phi(V_{e})} \rho(r) dr }{
\int_{-\infty}^{\phi(V_{b})} \rho(r) dr},
 $$
where ${N_{sp}}^{(b)}$ and ${N_{sp}}^{(e)}$ are diversities 
at the beginning and the end of the decreasing interval, 
and $V_{e}, V_{b}$ are equilibrium value of $v$ at
the beginning and the end of the decreasing interval. 
For all the period of evolution  one has 
 \begin{equation} \label{Eqmu}
 N_{sp}^{(f)} \approx  N_{sp}(0) \Big(\int_{-\infty}^{\phi(V_{min})} \rho(r) dr \Big) \Big(\int_{-\infty}^{\phi(V_{0})} \rho(r) dr \Big)^{-1},
 \end{equation}
where 
$N_{sp}^{(f)}$  is a final diversity value and $V_{0}$ is the initial equilibrium
value of $v$.

Under certain assumptions, 
this formula can be generalized for the multi-resource case.
Again, let $\phi_i(v)=\phi(v)$ be the same for all species, but
mortalities $r_i$ could be different. The main additional assumption is that turnovers $D_i >>1$. Then one can show
that the attractor of system (\ref{HX2}), (\ref{HV2}) is a stable 
equlibria, and equilibrium value of $v_i$ are close to the corresponding  resource supplies $S_i$. Then, by the same arguments,  we obtain  
\begin{equation} \label{EqmuM}
 N_{sp}^{(f)} \approx  N_{sp}(0)\Big( \int_{-\infty}^{\phi_{min}} \rho(r) dr \Big) \Big(\int_{-\infty}^{\phi(S(0))} \rho(r) dr \Big)^{-1},
 \end{equation}
where $S=(S_1, S_2, ..., S_M)$ is the vector of resource supplies,
$S(0)$ is an initial value of that vector and
$
\phi_{min} 
$
is the minimal value of $\phi(S)$ on the whole evolution time interval. 
In other situations the problem is much  more complicated, and it will be considered in   future studies.

\section{Simulations for simplest model } \label{numres}

Numerical simulations are made for the simplest model
with a single resource considered in the previous section. We use the formula  (\ref{eqstatev}).  We suppose that at initial time moment we have  $50$ coexisting species and the parameters are
$N_{pool}=50$, $b_i=2$, $K_i=4$, where $i=1, ..., N_{pool}$, $D=5$,  and $S=10$.  The mortalities $r_i$ are independent random numbers distributed normally according to ${\bf N}(1, 0.1)$. The matrix $\gamma_{ij}$
is diagonal with entries $\gamma_{ii}=1$. To assess the species model characteristic we considered the perturbations of $\tilde S(q(\tau))$ defined by different models: the periodic and quasiperiodic one defined by
(\ref{quasitrend}), a purely random model with $q(\tau)=B(\tau)$ and $B(t)$ is a white noise, 
the model exhibiting the Kramers transitions from Subsection \ref{Q3} and the chaotic forcing (\ref{ENSO}). The results of simulations are presented in Figures \ref{noise} and \ref{noisy2}. 

To compare our results,  we consider perturbations of 
$\tilde S$ of the same amplitude normalized  as follows:
$$
\tilde S(q(\tau))=\epsilon (q(\tau)-\bar q)/var(q),
$$
where $\bar q$ is the mean over trajectories $q(\tau)$, and
$var(q)=\max q(\tau)- \min q(\tau)$. We have made
$50$ tests for each perturbation with random mortalities
and random $q(t)$, where $\epsilon=5$ and $\bar S=10$.
Using the test simulation ensemble we compute the number of extinctions $n_{ext}$, the number $N_{surv}$ of finally survived species, the mean size of extinctions $\bar N_{ext}$ and maximal size $\bar N_{ext, max}$ of extinctions.

%\begin{figure}[!h]
\begin{figure} 
  \centering
  \begin{minipage}[b]{0.45\textwidth}
    \includegraphics[width=\textwidth]{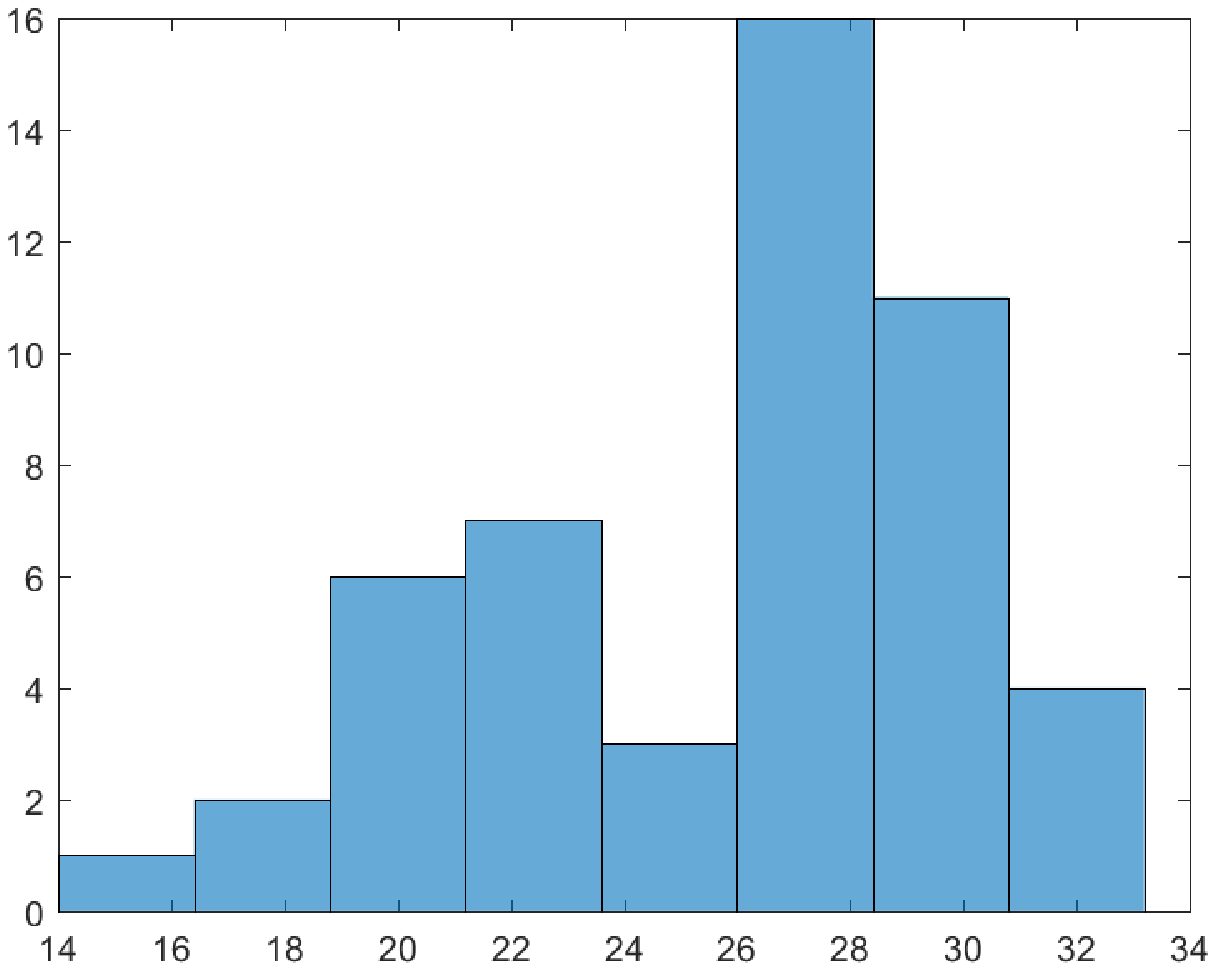}
    \subcaption {}
        \end{minipage}
 \begin{minipage}[b]{0.45\textwidth}
    \includegraphics[width=\textwidth]{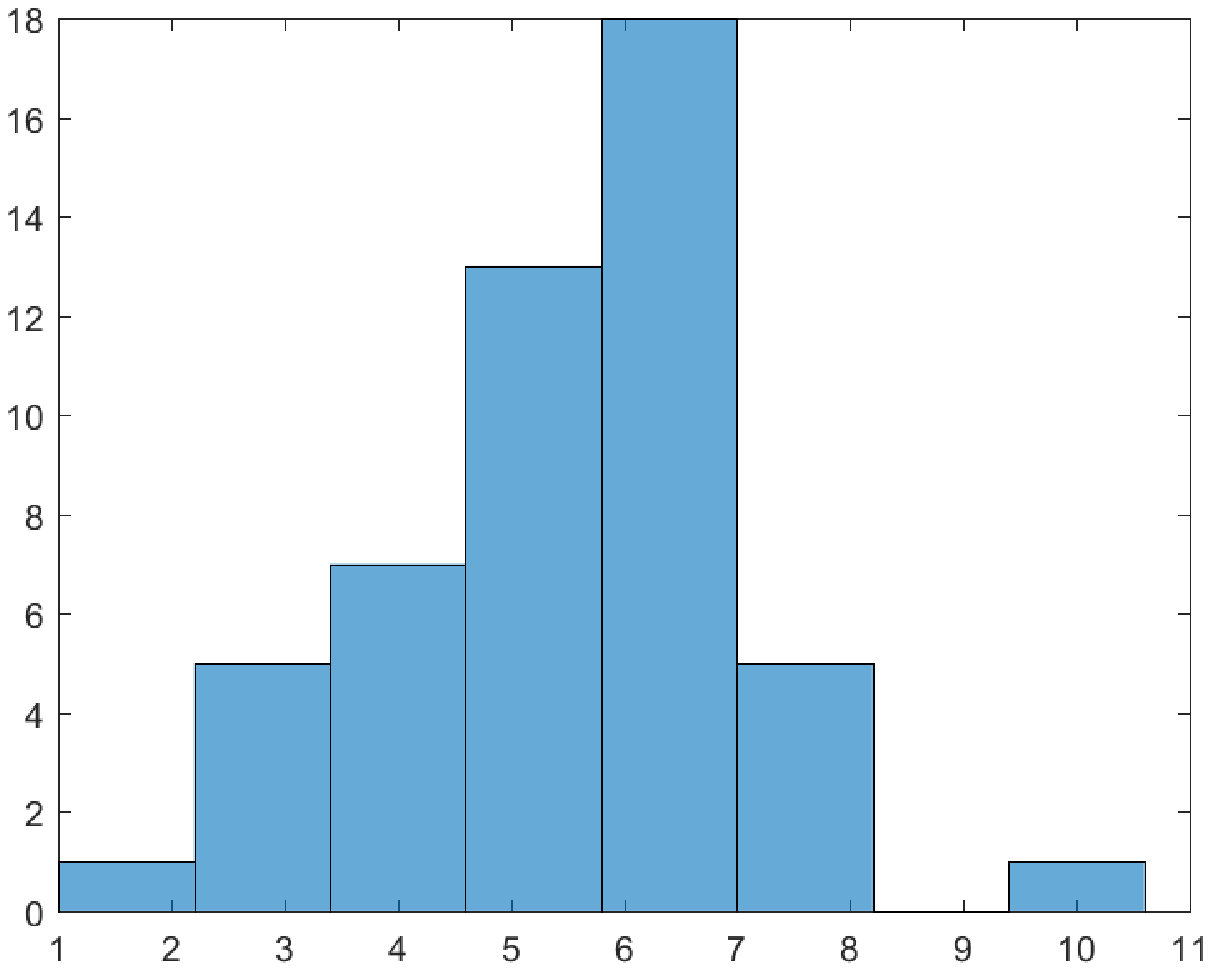}
    \subcaption {}
    \end{minipage}   
  %\begin{figure}[!tbp]  
\begin{minipage}[b]{0.45\textwidth}
    \includegraphics[width=\textwidth]{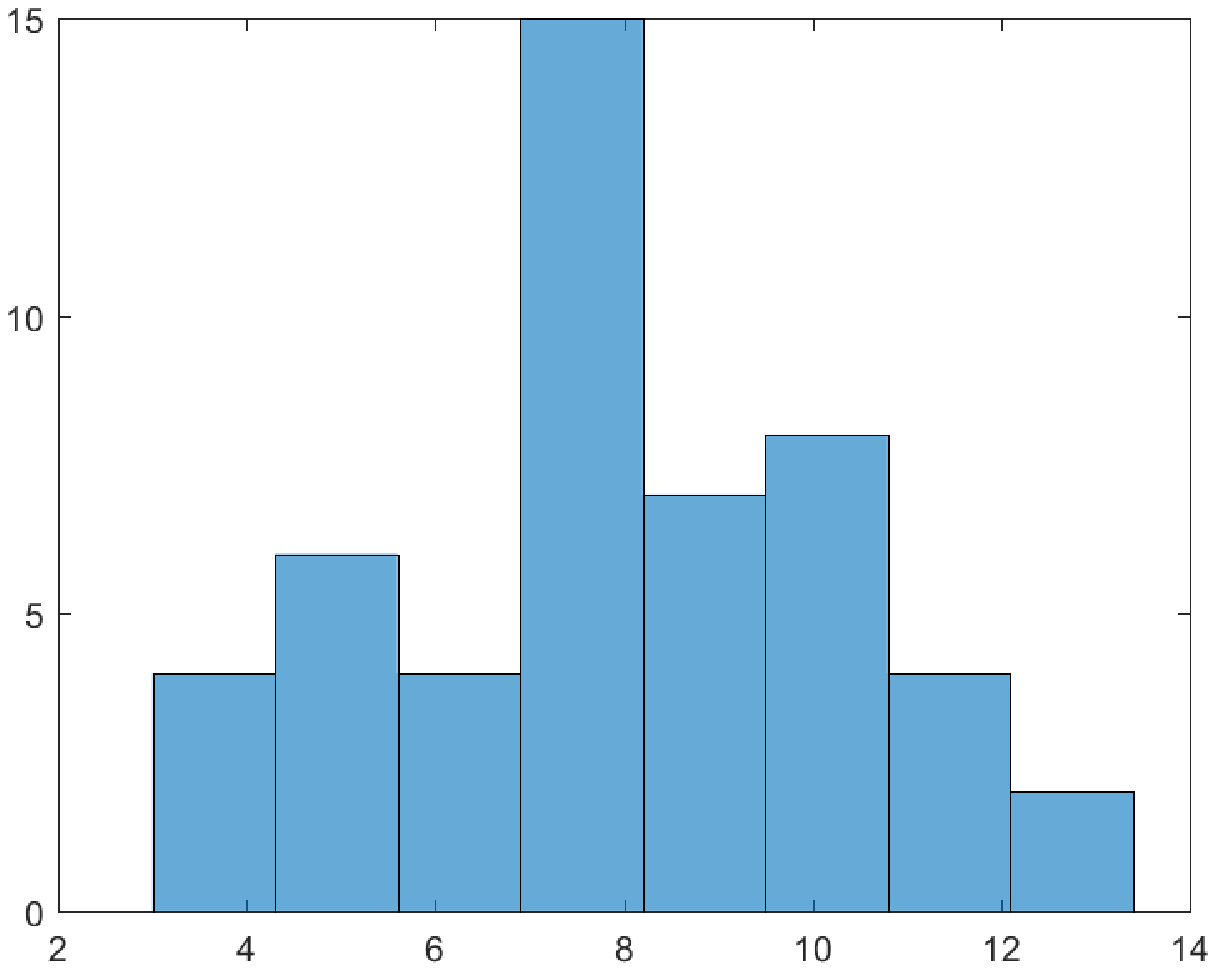}
    \subcaption {}
       \end{minipage}
\begin{minipage}[b]{0.45\textwidth}
    \includegraphics[width=\textwidth]{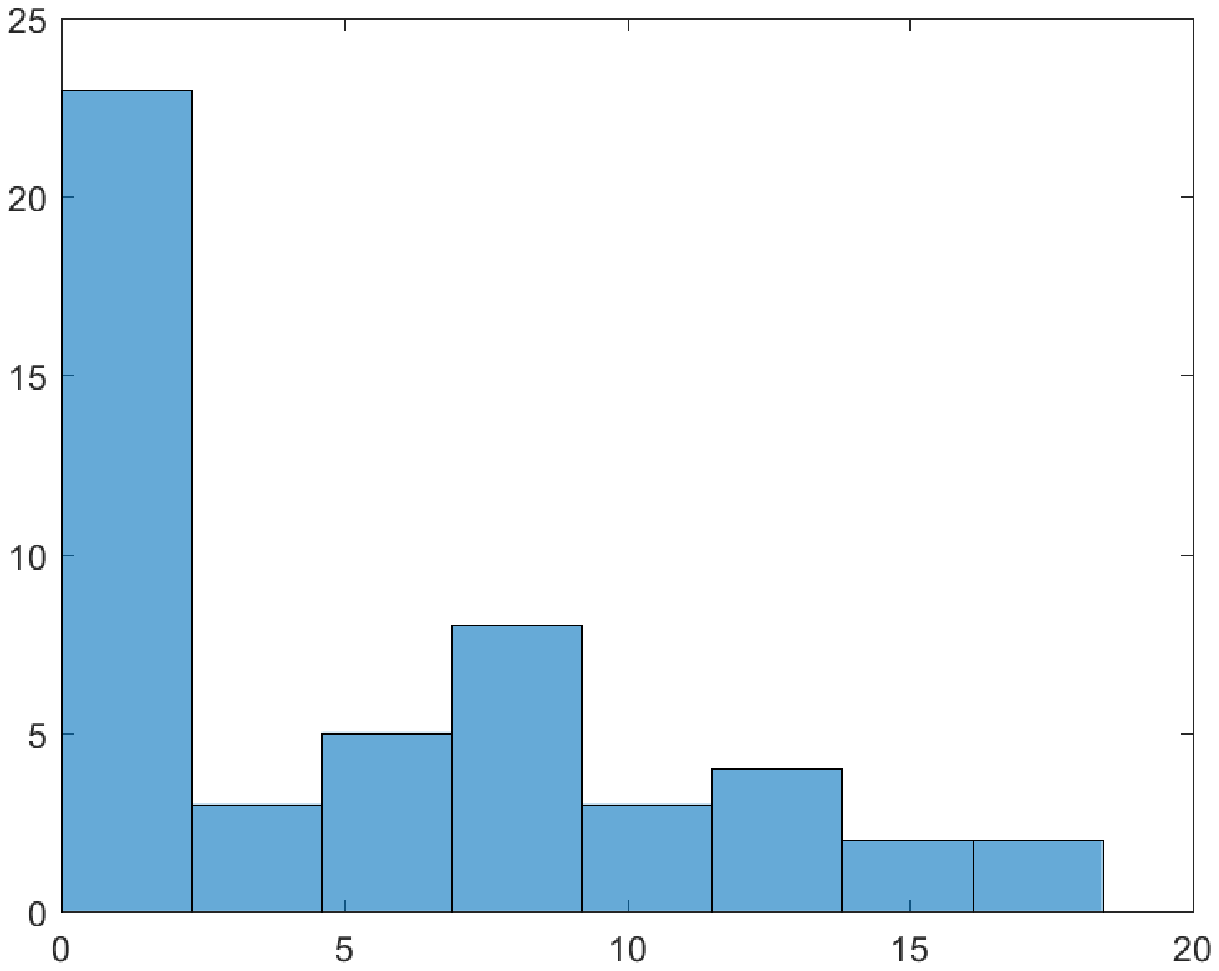}
    \subcaption {}
    \end{minipage}
    % \caption{\small This plot shows the histogram for the number     of extinctions within the period evolution $15000$ steps     for periodic perturbations \ref{quasitrend} with two different small frequencies $\omega_1=0.01, \omega_2=0.03$, where $c_1=1, c_2=0.5$. This simulation is made by the relation (\ref{eqstatev}).     Parameters are the same as on the previous Fig.     \ref{noise} (the same for all histograms).}
\caption{\small(a) The histogram for the number of extinctions within the period evolution $15000$ steps for periodic perturbations (\ref{quasitrend}) with two different small frequencies $\omega_1=0.01, \omega_2=0.03$ and with $c_1=1, c_2=0.5$ (simulation uses (\ref{eqstatev}) and parameters here for all plots are as for Fig. \ref{noise}). (b) This plot shows the histogram for the number
    of extinctions within the period evolution $15000$ steps
    under a random white noise perturbation. (c) This plot shows the histogram for the number     of extinctions within the period evolution $15000$ steps under a perturbation (Subsection (\ref{Q3})), where $a=1$, and it is perturbed by random uniformly distributed outliers of frequency     $p_{out}=0.1 $ so that system for $q$ exhibits the rare Kramers transitions from $q_1=\pm 1$ to $q_2=-q_1$. (d) This plot shows the histogram for the number     of extinctions within the period evolution $15000$ steps     for perturbations (\ref{ENSO}) with the frequency $\omega=0.01$ and $a= 0.1, c=0.2$. } 
    %\caption{\small  }
    %\caption{\small }
     \label{hyst}
    \end{figure}

%\end{figure}

Results for the number of extinctions are shown on Figure \ref{hyst}.
The distributions of the number of extinctions  $n_{ext}$ are essentially different for different cases. For the plots (a) and (b) on Figure \ref{hyst} it was checked by the Kolmogorov-Smirnov test. 
For the mean extinction size   $\bar N_{ext}$ we obtain the following. In quasiperiodical case $\bar N_{ext}\approx 1$,
for purely random perturbations $\bar N_{ext}\approx 2.7$,
for model from Subsection \ref{Q3}  $\bar N_{ext}\approx 2.0$, and
for the model (\ref{ENSO}) one has $\bar N_{ext}\approx 3.0$.
 By the end of evolution ($15000$ steps)
 $14$ species survived in the periodical case, $17$ in purely random case, and $14$ species survived for the model (\ref{ENSO}) and $16$ for the model with the Kramers transitions. We can conclude that the number of survived species on a long period depends on average intensity of environment perturbations but
 statistics and the size of extinctions depend on perturbation type. The periodic and chaotic forcing pertubation types, associated with periodic attractor and Gaussian ensemble eigenmode interaction type \cite{Bruun,Bruun2019} giving lower species survival. This type of interacting eigenmode environment perturbation may act to constrain the biota diversity accordingly.

\section{Conclusions} \label{Conc}
In this paper, using a fairly general niche model and the large deviations theory of stochastic processes, we show that there are possible three fundamentally different extinction scenario types that can be distinguished based on a state of the environment. To demonstrate the applicability of the proposed ideas we employ a resource model that describes a simple and easily understandable mechanism for resource competition in a population and takes into account species self-regulation, extinctions, forcing of the environment and time dependence of resources.

%In this paper, we study a resource model that describes a simple and easily understandable mechanism for resource competition in a population and takes into account species self-regulation, extinctions, forcing of the environment and time dependence of resources. %By the large deviations theory of stochastic processes, we show that there are possible three fundamentally different extinction scenario types that can be distinguished based on a state of the environment. 

The main result is that chaotic forcing perturbations can essentially affect biodiversity through a hysteresis effect for species extinctions. Different types of noise are considered, and it is found that the extinction probability sharply depends on the form of a noisy process. 
%Our results lead to the conclusion that a chaotically varying environment only affects a population when the averaged resource level is not large enough. Then the resource rate decrease can drive to a sharp biodiversity decline. However, that effect depends, in a non-trivial manner, on the type of environmental forcing. 
We also show that some environmental fluctuations may lead to dramatic consequences even if an averaged resource supply is sufficient to support population survival. In this case, the population can be destroyed by environmental noises. It is important to establish how many local attractors are generated by the internal wave interactions of the dynamical systems that defines environmental forcing and the location of these attractors with respect to critical level sets for resource supply. The periodic and chaotic forcing systems used here can exhibit eigenmode level repulsion that places the system in a chaotic class known as a Gaussian ensemble \cite{Bruun2019}.

This work contributes to the statistical physics of the ecological niche theory. The key works in this area \cite{Kes15, Dick16, Tix} demonstrate that stochastic processes can induce phases transitions in population from a niche phase where species competitions define the dynamics of the system to a neutral phase where stochasticity is the main driver of the population dynamic.  Our results show that the behavior of niche and neutral models is quite different when we take into account the environmental fluctuations.  

In addition, our results can be interesting for the biodiversity problem.  A globally prevalent generalist species like plankton may benefit from the effects of periodic and chaotic environmental forcing \cite {Cornwell2018} which enables it to adapt and out-compete other species by providing and ecosystem resilience when faced with system state changes.  We also think that our results will be applicable to the studies of past mass extinctions because recently such extinctions are considering as phases of a natural, 'meta-evolution' quasi-cycle where the timing and magnitude of mass extinctions are essentially stochastic events \cite{Rom19}. %that can be linked to some fundamental nonlinear phenomena such as, for instance, the self-organized criticality \cite{Sol97} and to nonlinear dynamics and critical phenomena more generally \cite{Rot19}.

\section*{Acknowledgments} 
We would like to thank Prof. V. Kozlov, Prof. U. Wennergren and Prof. V. Tkachev (Linkoping University) for useful
discussion. We thank the Statistical and Applied Mathematical Sciences Institute (SAMSI) and the Mathematical Biosciences Institute (MBI) for their support of this work. JTB also thanks Louise Cornwell (PhD candidate at Plymouth Marine Laboratory, UK) for useful discussions on plankton resilience and oceanic measurement.
The authors are grateful for financial support from the Government of the Russian Federation through the Mega-grant No.074-U01. We also acknowledge support from the the Russian Foundation for Basic Research (RFBR) under the Grants No. 13-01-90701 mol\_rf\_nr, No.16-34-00733 mol\_a\, and No.16-31-60070 mol\_a\_dk. In addition, we gratefully acknowledge support from the Division of Mathematical Sciences at the U.S. National Science Foundation (NSF) through Grant No. DMS-1743497. JTB also gratefully acknowledge the UK Research Councils funded Models2Decisions grant (M2DPP035: EP/P01677411), ReCICLE (NE/M00412011) and Newton Funded China Services Partnership (CSSP grant: DN321519) which helped fund this research.

\section*{Appendix}
\label{Appendix}

Here we outline the Freidlin-Wentzell theory.
Let the set of all possible trajectories $p(t) \ t \in [0, T]$  
defined on the time interval $[0, T]$ be equipped by the standard norm $||\cdot||_{\infty}$.
Then the set of trajectories with bounded norm $||\cdot||_{\infty}$ 
becomes a Banach space, which  will be denoted  ${\mathcal B}_T$.

Following \cite{FW} we define the rate function $I(q(\cdot))$, defined on the set of the trajectories $q(\cdot)$ by
\begin{equation} 
I(q(\cdot))=\frac{1}{2} \int_0^T |\frac{dq(t)}{dt} - Q(q(t)|^2 dt.
\end{equation}
For each closed subset ${\mathcal S} \subset {\mathcal B}_T$ of trajectories  let us consider 
the quantity
$$
P_{\mathcal S} =\inf_{q \in {\mathcal S}} I(q(\cdot)) 
$$
where we take the infimum over all possible trajectories belonging to  ${\mathcal B}_T$, which lie in the set
${\mathcal S}$ for each $t \in [0, T]$.
Then, according to \cite{FW} one has
\begin{equation} 
\liminf_{\epsilon \to 0} \epsilon \ln Prob\{q \in {\mathcal S}  \}  \le - P_{\mathcal S}
\end{equation}
and
\begin{equation} 
\limsup_{\epsilon \to 0} \epsilon \ln Prob\{q \in {\mathcal S}  \} \ge -P_{\mathcal S}.
\end{equation}

Let us define the distance $d(q, q')$ between two points
$q$ and $q'$  by
$$
d_{FW}(q, q')=\frac{1}{2}\inf_{p(\cdot) \in B(q, q')} I(p(\cdot)).
$$
where we take the infimum over the set $B(q, q')$ of the trajectories
$p(t)$ such that $p(0)=q$ and
$p(T)=q'$ and over all $T>0$.

The distance between the two sets $A$ and $B$ is defined as
$dist(A, B)= \inf_{q \in A, q' \in B} d_{FW}(q, q')$. 
The main property of $dist(q, q')$, needed for us,  we use is as follows.

The probability $P_{c, \epsilon}$ to attain the critical value starting from a point on a local  attractor ${\mathcal A}_Q$ satisfies the estimate 
\begin{equation} \label{probest}
\lim_{\epsilon \to 0} \epsilon \log P_{j, \epsilon} = -  \inf_{q \in {\mathcal A}_Q, q' \in {\mathcal O}({\Delta S}_j)} d_{FW} (q, q').
\end{equation}
By the definition of $dist(q, q')$ it is easy to show
 that if $q$ is a starting point of a trajectory of our dynamical system leading to $q'$ then $dist(q, q')=0$. 
Therefore, if the points
$q$ and $q'$ lie in the same connected component ${\mathcal A}_Q$ of the attractor then $d_{FW}(q, q')=0$.  However, if $q$ is in
a connected component of the attractor and $q'$ lie outside that component, then $d_{FW}(q, q')>0$.

%An unnumbered section, e.g.\ \verb"\section*{Funding}", may be used for grant details, etc.\ if required and included \emph{in the non-anonymous version} before any Notes or References.
%\section*{Notes on contributor(s)}
%An unnumbered section, e.g.\ \verb"\section*{Notes on contributors}", may be included \emph{in the non-anonymous version} if required. A photograph may be added if requested.
%\section*{Nomenclature/Notation}
%An unnumbered section, e.g.\ \verb"\section*{Nomenclature}" (or \verb"\section*{Notation}"), may be included if required, before any Notes or References.
%\section*{Notes}
%An unnumbered `Notes' section may be included before the References (if using the \verb"endnotes" package, use the command \verb"\theendnotes" where the notes are to appear, instead of creating a \verb"\section*").

\bibliographystyle{plain}

\end{document}